\documentclass[galaxies,review,accept,oneauthor,pdftex]{Definitions/mdpi} 
\newcommand{\aap}{Astron. Astrophys.}
\newcommand{\aapr}{Astron. Astrophys. Rev.}
\newcommand{\actaa}{Acta Astron.}
\newcommand{\aj}{Astron. J.}
\newcommand{\apj}{Astrophys. J.}
\newcommand{\apjl}{Astrophys. J. Lett.}

\newcommand{\mnras}{Mon. Not. R. Astron. Soc.}
\newcommand{\pasj}{Publ. Astron. Soc. Jpn.}
\newcommand{\pasp}{Publ. Astron. Soc. Pac.}
\newcommand{\pasa}{Publ. Astron. Soc. Aust.}

\firstpage{1} 
\pubvolume{9}
\issuenum{2}
\articlenumber{28}
\pubyear{2021}
\copyrightyear{2021}
\externaleditor{Academic Editor: {Roberto Mignani}} 
\datereceived{1 March 2021} 
\dateaccepted{13 April 2021} 
\datepublished{28 April 2021} 
\hreflink{https://doi.org/10.3390/\linebreak galaxies9020028} 

\Title{{Eclipsing} Systems {with Pulsating Components (Types $\beta$ Cep, $\delta$ Sct, $\gamma$ Dor or Red Giant)} in the Era of High-Accuracy Space Data}

\TitleCitation{Eclipsing Systems {with Pulsating Components (of Type $\beta$ Cep, $\delta$ Sct, $\gamma$ Dor or Red Giant)} in the Era of High-Accuracy Space Data}

\Author{{Patricia Lampens} \orcidA{}}

\AuthorNames{Patricia Lampens}

\AuthorCitation{Lampens, P.}

\address[1]{%
Royal Observatory of {Belgium}, Ringlaan 3, 1180 Brussels, Belgium; patricia.lampens@oma.be}

\abstract{Eclipsing systems are essential objects for understanding the properties of stars and stellar systems. 
Eclipsing systems with pulsating components are furthermore advantageous because they provide accurate constraints on the component properties, as well as {a complementary method for pulsation mode determination, crucial} for precise asteroseismology.
The outcome of space missions aiming at delivering high-accuracy light curves for many thousands of stars in search of planetary systems has also generated new insights in the field of variable stars and revived the interest of binary systems in general. The detection of eclipsing systems with pulsating components has particularly benefitted from this, and progress in this field is growing fast. {{In this review}}, 
we showcase some of the recent results obtained from studies of eclipsing systems with pulsating components based on data acquired by the space missions {\it Kepler} or TESS. {We consider different system configurations including semi-detached eclipsing binaries in (near-)circular orbits, a (near-)circular and non-synchronized eclipsing binary with a chemically peculiar component, eclipsing binaries showing the heartbeat phenomenon, as well as detached, eccentric double-lined systems. All display one or more pulsating component(s). Among the great variety of known classes of pulsating stars, we discuss unevolved or slightly evolved pulsators of spectral type B, A or F and red giants with solar-like oscillations. Some systems exhibit additional phenomena such as tidal effects, angular momentum transfer, (occasional) mass transfer between the components and/or magnetic activity.  How these phenomena and the orbital changes affect the different types of pulsations excited in one or more components, offers a new window of opportunity to better understand the physics of pulsations. }}


\begin{document}


\section{Introduction}\label{sec:1}

{Eclipsing systems are essential objects for understanding the properties of stars and stellar systems in general. Without these celestial objects, we would not have direct and {accurate} (relative and absolute) measurements of the radii and masses for a variety of stars, both nearby and further away, to confront with theoretical models of stellar structure and evolution~\cite{2010A&ARv..18...67T}. It would be arduous to measure stellar {properties} such as variations in surface brightness (limb and gravity darkening, reflection effect, \ldots), tidal or ellipsoidal flattening, surface inclinations without the possibility to observe the changes on the stellar surfaces during the eclipse phases {and be able to model these parameters adequately}~\cite{2018maeb.book.....P}. Eclipsing binary (EB) systems with pulsating components are furthermore advantageous because they provide accurate constraints on the component properties, as well as {a complementary method for pulsation mode determination (during the eclipses), crucial} for precise asteroseismology. In addition, they may undergo a series of phenomena which are due to the (possibly strong) gravitational forces acting on the components, which in turn can influence the stellar pulsations. }
\par EB systems can be classified as detached (D), semi-detached (SD) or (over)contact systems (C) according to the filling properties of their components' Roche lobes~\cite{1955AnAp...18..379K}. {Concurrently,} many different types of pulsating stars exist ranging from the long-period red giant variable stars to the extremely short-period B-type pulsating subdwarfs and white dwarfs (for a schematic overview encompassing 15 classes of pulsators, cf. Figure~1 in~\cite{2019A&A...623A.110G} which is an updated version of the variability tree presented by \citet{2008JPhCS.118a2010E}).
In addition, {most} types of pulsators {display} a large variety of pulsational properties that can be explored in combination with the different system configurations.~To maintain the focus on the {kinds of} interactions that can occur between the gravitational forces in the systems {on the one hand}  and the pulsations of their constituents {on the other hand}, we restricted our article selection to EB systems with either a main-sequence or slightly evolved pulsating component of spectral type B, A or F, or {with} a pulsating red giant star. 
\par Among the pulsator types, we pay specific attention to $\delta$ Scuti ($\delta$ Sct) stars, which are Population I stars of spectral types A-F with short-period, low-radial order pressure ($p$) modes, $\gamma$ Doradus ($\gamma$ Dor) stars, which are stars of {similar} spectral types exhibiting high-radial order gravity ($g$) modes with periods from 0.3 to 3~d--as well as the hybrid pulsators {among these classes} {--}, 
$\beta$ Cephei  ($\beta$ Cep) stars, which are late-O to early-B type non-supergiant stars with periods from 0.1 to 0.6 d, and red giant stars (RGs) exhibiting global solar-like oscillations 
\cite{2008JPhCS.118a2010E}. In contrast to EB systems with $\delta$ Sct-type pulsating components whose numbers are steeply rising mostly thanks to the space missions~\cite{2017MNRAS.465.1181L}, EB systems with massive $\beta$ Cep-type pulsating components are more rarely detected~\cite{2020MNRAS.497L..19S}. EBs with main-sequence pulsating companions have been discovered in systems of varying configurations, whereas EBs with solar-like pulsating RGs generally belong to well detached systems where the components are not expected to be strongly influenced by tidal effects and/or mass transfer~\cite{2013ApJ...767...82G}.~For these reasons, there is a larger diversity of interactions that has recently been reported in EBs with pulsating components of the intermediate spectral types A--F. 
\par Since the first ground-based discoveries, space missions with planet-finding objectives such as {\it CoRoT}~\cite{2006ESASP1306...33B}, {\it Kepler}~\cite{2010EGUGA..12.3890B, 2010PASP..122..131G} K2~\cite{2014PASP..126..398H}, TESS~\cite{2015JATIS...1a4003R} etc. have provided us with a bounty of newly detected eclipsing (binary and multiple) systems with oscillating components allowing for a rapid increase of {in-depth} studies. 
Such systems are very challenging objects but they offer great opportunities to study the pulsations in a close {system} configuration, where the tidal effects and/or mass transfer {events} can perturb, excite or damp the pulsations. This is why, {on top of providing an optimal characterization of their components through the determination of accurate fundamental stellar parameters based on high-quality data and multi-data modelling,} they are considered to be particularly interesting targets for~{asteroseismology}. 

\subsection{EBs with Main-Sequence (or Slightly Evolved) A/F-Type Pulsators}\label{sec:1.1}

\textls[-10]{Except for AB~Cas and Y~Cam, two Algol-type EBs of orbital periods 1.3668~d~\cite{1998A&A...340..196R} and 3.3055~d~\cite{2002A&A...391..213K}, respectively, including a $\delta$ Sct pulsating component, the first systematic detections of main-sequence A/F-type pulsating components of EB systems of type Algol (EA) from the ground go back to 2002~\cite{2002ASPC..259...96M, 2002IBVS.5314....1K, 2002IBVS.5325....1K}. \citet{2005ASPC..333..197M} presented a list of 20 known oscillating, mass-accreting components of SD Algol-type eclipsing binaries (that he named ``oEA stars'' for oscillating Algol-type eclipsing binary stars). Such systems (still) experience on-going mass transfer via the inner Lagrange L1 point leading to (epochs of) mass accretion unto the atmosphere of the pulsating component. Mass accretion introduces changes of the radius, mass and density as well as {evolutionary changes} of the pulsating star which depend on the mass {transfer} rate~\citep{2005ASPC..333..197M}. Over 70 oEA stars are presently known~\cite{2017MNRAS.465.1181L}. 
\par Other types of EBs may contain main-sequence A/F-type pulsating components as well. V392~Ori is a case of a SD EB of type $\beta$ Lyrae which belongs to the group of low-mass binary systems in the evolutionary state just off the main sequence. \citet{2002PASJ...54..139N} argued that V392~Ori is in an active state (with a magnetically active region on the surface of the primary) and probably {in a phase of mass transfer between its components}. \citet{2015AJ....150...37Z} revealed an additional signal in the residuals of the binary light curve with a periodicity of $\sim$0.0246~d consistently detected in the BVRI bands. The derived pulsation constant and the pulsational-to-orbital period ratio indicate that the primary component is of type $\delta$ Sct. {A catalogue containing 92 $\delta$\,Sct stars in EBs was presented by \citet{2017MNRAS.470..915K}}.}

\subsection{EBs with Main-Sequence (or Slightly Evolved) B-Type Pulsators}\label{sec:1.2}

{In contrast to the number of EBs known to include a $\delta$ Sct and/or $\gamma$ Dor pulsator, the number of EBs known to include a $\beta$ Cep pulsator is much smaller. The first $\beta$ Cep pulsating star belonging to an EB system was detected by \citet{1978IBVS.1508....1J}. In a review on intrinsic variability among binary and multiple systems, \citet{2006ASPC..349..137P} mentioned four $\beta$\,Cep pulsating stars in EB systems known at the time: 16 Lac, V381 Car, $\eta$~Ori, and $\lambda$~Sco. 16 Lac  is a well detached system with a near-circular orbit of period 12.0968 d~\cite{2015MNRAS.454..724J}. V381 Car is a detached system with an an almost circular orbit of period 8.32457 d, probably without apsidal motion~\cite{2005A&A...429..631F}. $\eta$~Ori is a quintuple system with two eclipsing early-B type components revolving in an inner orbit of period 7.989 d~\cite{1996A&A...310..164D}. \mbox{\citet{1996A&A...310..164D}} revealed line profile variability for the Ab component, which they interpreted as pulsations of type $\beta$ Cep, though the controversy about the nature of the short-period variability remains. $\lambda$~Sco~\cite{2006MNRAS.370..884T} is a hierarchical triple system with an inner pair of B-type stars in an orbit of period \mbox{5.9525 d} and  eccentricity 0.26, for which the precision on the masses of the components is of the order of 12\%. V916 Cen (with an orbital period of 1.46323 d), HD~101838 (with an orbital period of 5.41167 d), V4386 Sgr (in a near-circular orbit of period 10.79824 d), and HD~168050 (with an orbital period of 5.02335 d) are all detached systems~\cite{2010AN....331.1077D}. Most of these EBs are single-lined systems, for which a direct mass determination is not feasible. $\alpha$ Vir (Spica) is an ellipsoidal system in an eccentric orbit with an orbital period of 4.0145 d that was also observed by the satellite MOST~\cite{2016MNRAS.458.1964T}. The latest TESS discoveries represent a significant increase of new members, e.g., V453 Cyg~\cite{2020MNRAS.497L..19S}, VV Ori~\cite{2021MNRAS.501L..65S}, CW Cep~\cite{2021AJ....161...32L}.~The availability of the high-quality TESS light curves in combination with the component radial velocities of these double-lined EB systems is particularly helpful for reaching far better accuracy on the orbital parameters, therefore on the fundamental parameters of the (pulsating) components. In the case of V453 Cyg (cf. Section~\ref{sec:2.7}), the component masses were derived to a precision of 1.6\%. For VV Ori (with a circular orbit of period 1.485 d), the component masses are known to a precision of 6\%, while for CW Cephei (with a slightly eccentric orbit of period 2.729132 d and apsidal motion) they are known to better than 1\%. } 

\subsection{EBs with Solar-Like Oscillating RGs}\label{sec:1.3}

{KIC~8410637 is the first case of an oscillating RG belonging to an EB system discovered with {\it Kepler}~\cite{2010ApJ...713L.187H}. \citet{2013ApJ...767...82G} estimated the scientific potential of RGs in EBs based on the {\it Kepler} lists of 13,698 RGs (27 September 2011) and 2616 EBs. 
After having cross-correlated both lists, they found 12 new bona fide candidates in detached EB systems with orbital periods greater than 19 d, one in an ellipsoidal binary system with tidally induced oscillations, and 10 hierarchical triple systems with two low-mass components in a close, inner orbit and a pulsating RG star as the third component moving in an eccentric, outer orbit. Gaulme {et al.}\,~\cite{2013ApJ...767...82G,2014ApJ...785....5G,2016ApJ...832..121G} derived orbital and dynamical stellar parameters based on the modelling of the {\it Kepler} light curves and radial velocities for several samples of pulsating RGs in EBs. From a comparison between the masses and radii obtained from binary modelling with the values obtained from the asteroseismic scaling relations for 10~RGs in double-lined EBs, the authors concluded that the masses and radii based on the seismic scaling relations are systematically overestimated by about 15 and 5\%, respectively.} 

\subsection{Article Selection and Scope}\label{sec:1.4}

\textls[-15]{{{Our objective} is not to provide an exhaustive list of EBs with pulsating components of types $\beta$ Cep, $\delta$ Sct, $\gamma$ Dor or solar-like RGs that were detected and observed by the space missions.~Instead, we aim to provide a selective view {{on the new opportunities and insights}} 
that such studies can offer. Our selection consists in the following case studies: (a) SD EBs with  (near-)circular orbits and short orbital periods (Sections~\ref{sec:2.3},~\ref{sec:2.5}--\ref{sec:2.7}), (b) an ellipsoidal system with an intermediate orbital period (Section~\ref{sec:2.4}), (c) detached systems with eccentric orbits having both short and intermediate orbital periods (Sections~\ref{sec:2.1} and~\ref{sec:2.2}), (d) and well detached systems with eccentric orbits and long orbital periods (Section~\ref{sec:2.8}). 
\par Regarding the possible interactions between the gravitational forces in the systems and the pulsations, we will showcase tidally excited ({gravity} or $g$) modes 
\mbox{(Sections~\ref{sec:2.1} and~\ref{sec:2.2})}, tidally split ({pressure} or $p$, as well as $g$) modes (Sections~\ref{sec:2.3},~\ref{sec:2.6} and ~\ref{sec:2.7}), tidally perturbed modes (e.g., due to non-linear interactions, (Section~\ref{sec:2.1}), weak tidal influence (Section~\ref{sec:2.4}) and no (or untracable) tidal {{influence}} (Section~\ref{sec:2.8}). With respect to stellar interiors and surfaces, we may retrieve information about, e.g., the stellar density (Section~\ref{sec:2.1}), the core-to-surface differential rotation (Section~\ref{sec:2.4}), stellar evolution (Section~\ref{sec:2.8}) as well as the magnetic fields affecting the surfaces (Section~\ref{sec:2.3}) and the pulsations (Section~\ref{sec:2.5}).
{{In the following section, each case is presented in two parts: the first part highlights the results as presented by the authors, whereas the second part 
consists of a short summary followed by a discussion or~comment.}}} } 

\section{Analyses and Results}\label{2}
\vspace{3pt}

\subsection{KIC~4544587: An Eccentric, Short-Period Binary System with $\delta$ Sct Pulsations and Tidally Excited Modes}\label{sec:2.1}

\subsubsection{Results}
\citet{2013MNRAS.434..925H} analysed {\it Kepler} photometry and ground-based spectroscopy of KIC~4544587, a {\it Kepler} EB with an orbital period of 2.189~d and a notable eccentricity of 0.28. The light curves show two deep, unequal eclipses as well as a {{brightening}} in the shape of a heartbeat (at periastron). These data were used for binary modelling with {the modelling tool} PHOEBE~\cite{2011ascl.soft06002P} and a subsequent frequency search in an iterative way. The modelling includes {rapid} apsidal motion (at a rate of 182~yr/rotation cycle, possibly caused by tidal effects) and pseudo-synchronization for both components at periastron passage. The final {binary} model was derived from the light curve prewhitened for 31 pulsation frequencies. This model is not a perfect representation of the data, in particular for the primary and secondary eclipse phases. According to the authors, the discrepancy is due to a combination of (i) the presence of pulsations that are commensurate with the orbital period and (ii) the precision of the {\it Kepler} data {{illustrating the need for improved limb darkening} {{coefficients}} {and albedo's}}.  
\par {{From a frequency analysis of the residual data, they identified}} 31 modes, 14 in the $g$ mode region {(between 0 and 5 d$^{-1}$)} and 17 in the $p$ mode region {(between 30 and 50~d$^{-1}$)}. {Many extracted $p$ modes appear in multiplets split by the} orbital frequency. From the pulsation constants and the fundamental parameters, they determined that the $p$ modes correspond to radial overtones in the range 3 $\le$ $n$ $\le$ 5. Of the 14 identified $g$ modes, 8 were found to be tidally excited. 
\par {{\citet{2013MNRAS.434..925H}} {derived the atmospheric parameters} 
and the projected rotational velocities of the components from optimal fitting of the disentangled Balmer lines and the least blended metal lines, respectively}.~The effective temperatures (T$_{\rm eff}$) and log $g$ values were obtained by fitting the disentangled spectra with the light factors set as free parameters. Stellar evolution and pulsation models for single, spherically symmetric, non-rotating stars were adopted. The models for the primary star present few excited $p$ modes, while the models for the secondary {{show}} unstable $\delta$ Sct-type frequencies in a much wider range ({{in closer agreement with the observations}}). However, none of the considered models is able to produce unstable $g$ modes. On the other hand, the {{unexplained}} $g$ modes could also be tidally (i.e., non-linearly driven by the dynamical tide) or stochastically excited~(\cite{2013MNRAS.434..925H}, and references therein). 

\subsubsection{Conclusions--Discussion}
KIC~4544587 is a short-period EB with self-excited pressure ($p$) and gravity ($g$) modes, tidally excited ($g$) modes, as well as tidally perturbed ($p$) modes. Although the system is a short-period though eccentric binary, {{no sign of activity was reported and}} no process (episodes) of mass transfer was {considered} (as occurred in, e.g.,\,RZ~Cas~\cite{2018MNRAS.475.4745M}). In order to explain both the eccentricity and the {{fast}} apsidal motion, the authors suggest the presence of (extreme) tidal interactions or an (undetected) companion. Both components lie in the $\delta$ Sct instability strip, {{it is not sure at this stage which one is the $\delta$ Sct-type pulsator}}. The current models seem to rule out self-excited $g$ modes. 
{{Whether one of the components might undergo self-excited 
$g$ modes would need to be verified with improved models}}. {{If not, this study might reveal}} a $\delta$ Sct pulsator in an eccentric {binary} with {$g$ modes} also driven by the dynamical tide. 

\subsection{KIC~4142768: An Evolved $\gamma$ Dor-$\delta$ Sct Hybrid Pulsating EB with Tidally Excited~Oscillations}\label{sec:2.2}
\vspace{3pt}

\subsubsection{Results}
\citet{2019ApJ...885...46G} more recently presented a study of KIC~4142768, an eccentric EB system {of intermediate period 13.995802 d} consisting of two slightly evolved A-type stars. The {\it Kepler} light curve is that of a heartbeat star. Both components share the same origin in time and space (overall chemical composition and location) and have similar properties (i.e., atmospheric properties and mass). The only differences {are} the radius (one is a little larger and more evolved) and the {presence (or absence) of} pulsations (one is a hybrid pulsator while the other may present a few excited $g$ modes). The system is also a double-lined spectroscopic binary (SB), which enables a complete characterization of the components based on the combined modelling of the light and component radial velocity curves.
\par Most of the $g$ modes consist of self-excited modes of $\gamma$ Dor type (between 0.1 and 1.5~d$^{-1}$) and tidally excited modes driven by the dynamical tide (between 0.1 and \mbox{3.0 d$^{-1}$}). In the range between 1.4 and 2.0 d$^{-1}$, the peaks {show a near-equal period spacing} ($\Delta$P) of about {0.0347 d$^{-1}$} (3000~s). From this, assuming that the spin and orbital axes are aligned, the authors conclude that these are most probably prograde dipole sectoral modes (\mbox{$l$ = 1}, \mbox{$m$ = 1}). Fitting the P-$\Delta$P linear trend (using the asymptotic relation for high-order $g$ modes) suggests a near-core rotation rate of 0.006 d$^{-1}$. The $p$ modes, located between 15 and 18~d$^{-1}$, are shown to be consistent with an evolved stage which is near the end of the ZAMS. Further modelling seems to suggest that some of the remaining self-excited $g$ modes might be ($l$ = 2, $m$ = 2) modes. 
\par {{\citet{2019ApJ...885...46G} concluded that}} the primary star is likely a hybrid pulsator showing both $\delta$ Sct-type ($p$) modes and tidally excited and self-driven $\gamma$ Dor-type ($g$) modes, though some $g$ modes may also originate from the secondary component. The {hybrid pulsator in KIC~4142768 presents} a slowly rotating core and a sub-pseudo-synchronously rotating surface (\cite{2019ApJ...885...46G}, and references therein). 

\subsubsection{Conclusions--Discussion}
This study illustrates the advantages of studying an EB-SB system with {hybrid} pulsating components well. The full characterization allowed to determine the orbital properties of the system and the fundamental parameters, i.e., the component masses and radii, mass ratio, orbital inclination, and the eccentricity. {Using the spectra separation technique, the authors obtained} the component atmospheric properties: the effective temperatures, gravities, projected rotational velocities and the metallicity. A model was found giving a probable age of 1.0 $\pm$ 0.1~Gyr, with both components near the end stage of the ZAMS, just before the contraction phase. {The results allow a direct} confrontation between observations and theory with respect to various aspects.
\par On the one hand, using the codes MESA (stellar evolution~\cite{2011ApJS..192....3P}) and NADROT (stellar stability~\cite{1977AcA....27...95D}), {{the authors were able to match}} the observed $p$ modes and the large separation theoretically. On the other hand, using the codes MESA and GYRE (stellar pulsation~\cite{2013ascl.soft08010T}), {{they could match}} the tidally excited $g$ modes (i.e., specific orbital harmonics) theoretically. The slope and vertical displacement of the P-$\Delta$P diagram were used to provide information on the mode identification and the near-core rotation rate (assuming that the primary is the hybrid pulsator and spin-orbit alignment). With the surface rotational velocities {known}, an internal rotation profile was deduced. Given that both components have very similar characteristics, the origin of the pulsations {is intriguing}. Such information could provide even more constraints. Of course, the question why both components do not behave identically from a seismic point-of-view {remains}, i.e., which essential property distinguishes the companion from the (hybrid) pulsating primary? 

\subsection{The {\it Kepler} Eclipsing and Spotted Binary System KIC~6048106. Hybrid Pulsations and Tidal~Splitting}\label{sec:2.3}
\unskip
This study was performed in two steps. In a first paper~\cite{2018MNRAS.474.5549S}, an appropriate binary model was derived based on {\it Kepler} photometry and the modelling tool PHOEBE~\cite{2011ascl.soft06002P}. In the second paper~\cite{2018AcA....68..425S}, the residual light curve was used to perform a detailed frequency and pulsation analysis. 

\subsubsection{Results}
KIC~6048106 is a SD Algol-type eclipsing binary with F5-K3/4 components and an orbital period of 1.55936~d.~The system presents a (near-)circular orbit (e $\sim$ 0.01), slight asynchronous rotation and an orbital inclination of $\approx$ 73$^{\rm o}$. The light curve displays an {obvious} flux modulation observed during {both} quadrature phases.~Because of this modulation, the authors added a cold spot on the surface of the secondary component, a cool subgiant, into the model.~They derived the fundamental stellar parameters of each component based on a consistent binary model including a variable spot structure going from a low (L), through a medium (M) to a high (H) state of activity. They found a modulation period of 290 $\pm$ 7~d, which agrees very well with the {trend detected} in the {\it Kepler} Eclipse Time Variations of the secondary component.
These facts suggest the existence of a periodic cycle of (magnetic) activity at the surface of the secondary star~\cite{2018MNRAS.474.5549S}. 
\par A Fourier analysis of the residual light curve revealed significant frequencies in mainly three intervals: (i) between 1.9 and 2.9 d$^{-1}$, seven $g$ modes (with amplitudes {in the range} 200--1588 $\upmu$mag) were identified, (ii) between 7.4 and 15.2 d$^{-1}$ and between 19 and 22.5~d$^{-1}$, 34 $p$ modes (with amplitudes in the range 50--106 $\upmu$mag) were detected~\cite{2018AcA....68..425S}. 
\par {Among the low frequencies,} 1.9764~d$^{-1}$ is most dominant. Seven independent $g$ modes with a mean $\Delta$P of {0.0176 d} were found, six of which were already reported by \citet{2016ApJ...833..170L}.~The corresponding period--\'echelle diagram shows that the $g$ modes follow a single ridge with some wiggles. In the high-frequency region, two clusters of frequencies were detected. The most dominant frequencies of each cluster (excluding harmonics of the orbital frequency) are 11.745 and 20.960 d$^{-1}$. It should be noted that the maximum amplitude in the acoustic regime is about ten times smaller than that in the gravity regime. This is one of the reasons why \citet{2016ApJ...833..170L} did not detect them. Five regular multiplets with a mean {frequency} spacing of 0.64113 $\pm$ 0.00027 d$^{-1}$ ({in perfect agreement with}\mbox{ f$_{orb}$ = 0.64143 $\pm$ 0.00015 d$^{-1}$}) were identified, {including} 25 $p$ modes equidistantly split by the orbital frequency. The corresponding period--\'echelle diagram shows that all frequencies fit well into three ridges ({{possibly corresponding}} to $l$ = 0, 1 and 2). 

\subsubsection{Conclusions--Discussion}
This EA system {presents} a short-period, {almost} circular orbit. It consists of a {$\gamma$ Dor - $\delta$ Sct} hybrid pulsator (type F5) and a cool subgiant that shows a 290-d activity cycle (probably of magnetic origin). From the residual light curve, {{the authors extracted}} seven $g$ modes and 34 $p$ modes of much lower amplitude. {{They showed that}} most of the $p$ modes occur in multiplets equidistantly split by the orbital frequency. This is {one of the first studies} where the observations can be compared to the theory of stellar pulsations in close binary systems experiencing tidal forces. Since the equilibrium tide of a close binary system with a circular orbit leads to equidistant frequency splitting of an excited mode by the orbital frequency according to theoretical predictions~\cite{2003A&A...404.1051R, 2003A&A...409..677R, 2005ASPC..333...39S}, the frequency multiplets in the $p$-mode region can be interpreted as tidally split $p$ modes, similar to the conclusion of \citet{2016ApJ...826...69G} during the pulsation study of KIC~9851944. The question one might ask is whether the excited $g$ modes should ``feel'' the tidal distortion or how significant the impact of the equilibrium tide might be {in the near-core regions}.  

\subsection{KIC~10080943: An Eccentric Binary System Containing Two Pressure-  and Gravity-Mode Hybrid Pulsators}\label{sec:2.4}
\unskip
\citet{2015A&A...584A..35S} and \citet{2015MNRAS.454.1792K} presented a study of two hybrid pulsators in KIC~10080943, whereas \citet{2016A&A...592A.116S} performed a seismic modelling of the $g$ modes. Although this system is non-eclipsing, the goal and the applied methods are identical to the study of EB systems. For this reason and the diversity of results obtained, we decided to include it.

\subsubsection{Results}
KIC~10080943 is an ellipsoidal, double-lined and eccentric binary system with an orbital period of 15.3364~d and slowly rotating ‘twin stars' of spectral type F (both $v$.sin\,$i$'s are of the order of 10 km/s). Using PHOEBE~\cite{2011ascl.soft06002P}, the authors modelled the ellipsoidal variation and the reflection signal of the binary in the {\it Kepler} light curve. 
{From ground-based and phase-resolved, high-resolution spectra, the orbital parameters and the component spectra were determined using a spectral disentangling technique~\cite{2017ascl.soft05011I}.} Next, GSSP$\_$BINARY~\cite{2015A&A...581A.129T} was used to fit the disentangled component spectra. This code allows for a fitting of both spectra simultaneously and takes into account the wavelength dependence of the light ratio by replacing it with the ratio of the radii.
\par After removal of the best-fit binary model, the residual {\it Kepler} light curve was subjected to a Fourier analysis. In the high-frequency regime, the most prominent feature is the quintuplet around 17.3 d$^{-1}$, with a mean frequency splitting corresponding to twice the orbital frequency. In the triplet around 15 d$^{-1}$, a smaller splitting value of 0.1213 d$^{-1}$ was found. In the low-frequency regime, six period spacing series, which form rotationally split triplets of prograde, zonal and retrograde modes for one component, and doublets of prograde and retrograde modes for the other component were identified~\cite{2015MNRAS.454.1792K}. The detection of rotational splitting among the $g$ modes (sensitive to the near-core regions) as well as among the $p$ modes (sensitive to the outer regions) enabled them to also estimate the core-to-surface rotation rates for both stars. By computing the time delays of seven modes, i.e., the phase changes of the pulsation frequencies generated by the orbital motion~\cite{2014MNRAS.441.2515M}, the authors were able to associate the frequencies to either the primary or the secondary component. In this way, \citet{2015A&A...584A..35S} showed that this eccentric binary system contains two F-type hybrid pulsators of the $\gamma$ Dor-$\delta$ Sct type. 

\subsubsection{Conclusions--Discussion}
Although KIC~10080943 is a non-eclipsing system--which inhibits the determination of the absolute masses in a model-independent way{--}, 
it does provide hard constraints for a detailed modelling of its two pulsating components.~The {{physical association of the}} binary implies that the seismic properties of both {stellar} models must satisfy the conditions of equal age and composition, {as well as} the observed mass ratio.~Based on a fit of the observed mean $\Delta$P values with the corresponding asymptotic spacing values, {{\citet{2016A&A...592A.116S} showed that}} models with an age around 1.1~Gyr and a low core hydrogen mass fraction {{provide the best match}}. However, these models fail to predict the detailed period spacing structure. On the other hand, the authors found a coeval model with low masses and low metallicity that fits both the structure and the mean value of the period spacing. However, this model cannot reproduce the observed $p$-mode frequencies and is discrepant with respect to the fundamental parameters of the system. This shows both the complexity of such a detailed modelling and the weaknesses of the models. The amount of observational constraints for this system is very rich (e.g., fundamental and seismic parameters, near-core rotation rates, mean $\Delta$P values for two similar hybrid pulsators, origin of the pulsation modes), nevertheless it was not possible to find a model that simultaneously satisfies all the observational data.~Suggested reasons for this are (1) the limitations of 1D stellar models in terms of chemical mixing, (2) the use of a time-independent simplified convection theory for the thin stellar envelope and (3) the fact that tidal effects in the system were not accounted for (these forces were estimated to be small although the (orbit-dependent) rotational splitting of the $p$ modes may be a clue). Including these aspects in the theoretical framework (e.g., with MESA) may allow for an improved and successful modelling of this interesting system in the future~\cite{2021RvMP...93a5001A} . 

\subsection{HD~99458: First Time Ever Ap-Type Star as a $\delta$ Scuti Pulsator in a Short-Period Eclipsing~Binary?}\label{sec:2.5}
\vspace{3pt}

\subsubsection{Results}
\citet{2019MNRAS.487.4230S} reported the very first discovery of a pulsating Ap star in the EB HD~99458. The binary system consists of a chemically peculiar (CP) Ap-type star exhibiting $\delta$ Sct pulsations and a low-mass early-M type dwarf star that are on an inclined orbit with an orbital period of 2.722~d and grazing eclipses. 
\par The light curve comes from the K2 mission~\cite{2013MNRAS.434..925H}, while the high-resolution \'echelle spectra were collected at the Ond\v rejov observatory, Czech Republic, and Star\'a Lesn\'a observatory, Slovak Republic. The atmospheric properties of the primary component (\mbox{T$_{\rm eff}$ = 7600~K}, log~$g$ = 3.6, $\nu_{\rm mic}$ = 2~km/s) were derived as well as the abundances which are typical of a magnetic Ap star. The binary model was computed with PHOEBE~\cite{2011ascl.soft06002P} by fixing {already derived} parameters (mostly of the primary component) and assuming two cool spots on the surface of the primary component. The fundamental parameters of the secondary component, M$_{2}$ = 0.45 M$_\odot$ and R$_{2}$ = 0.59 R$_\odot$, were then obtained. Since the assumption of spin-orbit alignment contradicted the observed properties of the primary component, \citet{2019MNRAS.487.4230S} concluded that the rotational and orbital axes must be misaligned. On the other hand, the phased light curve indicates that the rotational velocity of the spotted primary star is synchronized with its orbital velocity. Therefore, although there is spin-orbit misalignment, the system is synchronized. 
\par From a frequency analysis of the light curve residuals, the authors identified at least six independent frequencies, with the dominant frequency at 19.2 d$^{-1}$ {and an amplitude of 1.28 ppt}. {Apart from short-periodic stellar pulsations, other options} were also considered (e.g., contamination by nearby stars, blends and granulation) but were rejected as possible explanations of the fast photometric variations. Thus, the CP primary component of HD~99458 exhibits pulsations of type $\delta$ Sct. 

\subsubsection{Conclusions--Discussion}
HD~99458 represents a special case of a short-period EB system with an anomalous $\delta$ Sct-type pulsator. Not only were there no magnetic Ap stars {known} in short-period binary systems until now, but the presence of a magnetic field (needed to explain the cool spots as well as the peculiar surface abundances) is rare among classical $\delta$ Sct stars~\cite{2017MNRAS.468L..46N}. Such a {unique} system offers the possibility to investigate {a variety of interactions: the interaction between binarity and magnetic fields, binarity and chemical composition,} {magnetism and pulsations}, and {to explore} how unstable atmospheric conditions (induced by the pulsations) may affect the mechanism of diffusion leading to an anomalous surface composition. 

\subsection{Discovery of Tidally Perturbed Pulsations in the Eclipsing Binary U~Gru: A Crucial System for Tidal Asteroseismology}\label{sec:2.6}
\vspace{3pt}

\subsubsection{Results}
\citet{2019ApJ...883L..26B} analysed the TESS light curve of U~Gru (TIC~147201138), an EA system {of period 1.88050~d} that (possibly) {experienced} mass transfer and exhibits low-amplitude pulsation modes {that are systematically offset with respect to the orbital harmonics. The primary component is of spectral type A5 and the system has a circular orbit.} Binary modelling and the totality of the primary eclipses confirms a near edge-on inclination, and indicates a possible (photometric) mass ratio of 0.17. The pulsation modes were extracted from the residual amplitude spectrum using iterative prewhitening and optimized using a multi-frequency non-linear least squares fit to the residual light curve. {In total, 22 frequencies were detected in the high-frequency range between 21 and 67 d$^{-1}$. Among these, 19 frequencies can be considered as part of a long series of frequencies with an average offset of 0.074 d$^{-1}$ from an orbital harmonic frequency, whereas the three remaining frequencies are differently offset from {other} orbital harmonics. 
\par This study revealed {several interesting} features: 
(i) a multiplet of low-amplitude $p$ modes {equidistantly} split by the orbital frequency; 
(ii) three independent $p$ modes which do not belong to any multiplet}; 
(iii) additional flux modulation in the light curve, particularly evident at the phases of ingress and egress of the primary eclipse;
(iv) an effect of amplitude magnification of the equidistantly split $p$ modes during the phase of primary eclipse. 

\subsubsection{Conclusions---Discussion}
U~Gru {is a potential new member of} the class of the oEA stars 
\cite{2005ASPC..333..197M}. {\citet{2019ApJ...883L..26B}} are acquiring phase-resolved high-resolution spectroscopy in order to accurately determine the best possible physical binary model. Although spots on the surface of a cool companion can explain flux variability occurring at specific phases in the light curve, {their} modulation usually {occurs} on a longer {time scale} when it is coupled to changes in (magnetic) activity, e.g., \citet[][]{2018MNRAS.474.5549S, 2019MNRAS.487.4230S}. 
The flux modulation {reported in the TESS light curve} 
{suggests} a faster process, {perhaps being} caused by {the presence} of accreted material. Indications for this may come from high-resolution spectroscopy 
\cite{2018A&A...615A.131L}. The series of tidally split $p$ modes could be caused by the equilibrium tide, {similar to the cases of KIC~9851944~\cite{2016ApJ...826...69G} and KIC~6048106~\cite{2018AcA....68..425S}}. The observation that the tidally split $p$ modes have the highest amplitudes during the phase of primary minimum is an interesting finding. 
It could be an amplification effect of the non-radial modes {during the eclipse}.
Ideally, this effect allows to resolve the pulsations across the surface of the pulsating star and {enables} identification of the non-radial modes, e.g.,~\cite{2005ASPC..333..258G}. With accurately determined physical properties of its components, this new member among the oEA stars {is a very promising object for studying the connection between tidal effects and $\delta$ Sct pulsations. } 

\subsection{Discovery of $\beta$ Cep Pulsations in the Eclipsing Binary V453 Cygni}\label{sec:2.7}
\vspace{3pt}

\subsubsection{Results}
{\citet{2020MNRAS.497L..19S} reported the discovery of $\beta$ Cep pulsations in V453~Cyg, an EB consisting of two slightly evolved, early B-type stars moving in a near-circular orbit with an orbital period of 3.89 d and an eccentricity of 0.025. The binary undergoes apsidal motion with a period of 72 years~\cite{1973A&A....25..157W}. The TESS light curve covering two months was binned according to phase and modelled with the Wilson–Devinney code (version of 2004)~\cite{1971ApJ...166..605W}, after removal of the strongest pulsation signal. A best-fit model was computed with fixed mass ratio, effective temperatures, rotation rates, and gravity darkening coefficients. 
\par The adopted radial velocity amplitudes and the component effective temperatures were based on the method of spectra disentangling~\cite{2018MNRAS.481.3129P}.~The authors determined the fundamental properties of both components, including precise masses and radii (measured with accuracies of 1.6\% and 0.6--1.0\%, respectively). They confirmed that the primary star is rotating almost synchronously with the orbit, whereas the secondary star is rotating faster by a factor of 1.4. Using discrete Fourier transform 
to calculate the amplitude spectrum, they extracted nine significant frequencies, two of which may be due to an imperfect binary model. The remaining significant frequencies represent low radial order $g$ and $p$ modes. Among these, six frequencies form a regular pattern with a frequency spacing equal to (an integer multiple of) the orbital frequency. The highest frequency is likely an independent $p$~mode. }

\subsubsection{Conclusions--Discussion}
{Using the TESS light curve and published spectroscopic results, the authors derived a physical model for this near-circularized SD EB system. They also detected a series of tidally split low frequencies (called 'tidally perturbed' frequencies in analogy with the study in Section~\ref{sec:2.6}, though with a different meaning than in Section~\ref{sec:2.1}), as well as an independent $p$ mode frequency.
\par This analysis is the starting point of a rare but important case, as the EB contains a massive pulsating component of type $\beta$ Cep.~The physical properties, including the masses and radii, are very accurately derived from the high-quality photometry coupled to recent spectroscopic results (adopting some fixed parameters), though not through a combined modelling approach where most parameters are simultaneously constrained. The mass ratio is approx. 0.8, which implies that both components could in principle be pulsating. Both components could possess distorted envelopes and produce a tidally split multiplet corresponding to a specific pulsation mode. It is important to explore how the low frequencies (the $g$ modes) are affected by the tides, since this has not been observed in some less massive oEA stars showing long series of consecutive tidally split $p$ modes (e.g., KIC 6048106 and U Gru). The single $p$ mode is apparently not affected, suggesting a radial mode, in which case this result would agree with standard theoretical prediction for pulsations in a close binary with equilibrium tides~\cite{2003A&A...409..677R}. Another {explanation would be} that the independent $p$ mode might correspond to a radial overtone originating from the secondary component (with a possible period ratio of 0.73). A refined discussion of the pulsations awaits more extensive spectroscopic data. Line profile variations were also reported in some previous {studies}. }  

\subsection{Oscillating Red Giants in Eclipsing Binary Systems: Empirical Reference Value for Asteroseismic Scaling Relation}\label{sec:2.8}
\vspace{3pt}

\subsubsection{Results}
{An in-depth study of three pulsating red giant stars (RGs) in eccentric, double-lined EBs was performed by \citet{2018MNRAS.478.4669T} based on high-resolution spectra and 4-year long {\it Kepler} light curves with the aim to test the validity of the asteroseismic scaling relations. These relations indeed assume that all stars have an internal structure homologous to the Sun. The RGs belong to detached EBs with relatively long orbital periods ranging from 175.4 to 987.4 days: KIC~8410637~\cite{2010ApJ...713L.187H}, KIC~5640750~\cite{2014IAUS..301..413G}, and KIC~9540226~\cite{2016ApJ...832..121G}. From the binary analysis, orbital solutions and the physical properties of the systems were obtained, leading to dynamically derived masses and radii. Individual component spectra for all three systems were extracted with a spectral disentangling technique~\cite{2017ascl.soft05011I}, from which the atmospheric component parameters such as the effective temperatures (T$_{\rm eff}$), metallicities and surface gravities of the RGs could be determined. On the other hand, the internal properties of oscillating RGs can also be inferred through the study of their global oscillation modes. From the analysis and a global modelling of the Fourier spectra of these oscillating RGs, the authors recovered the following set of seismic parameters, i.e., the frequency at maximum amplitude ($\nu_{max}$) and the large and small frequency separations.~They then computed the asteroseismic values of the stellar parameters (i.e., the masses, radii, mean densities, and surface gravities) by applying well-known asteroseismic scaling \mbox{relations~\cite{1995A&A...293...87K,2010ApJ...723.1607H}} as well as grid-based modelling. This enabled them to perform a thorough comparison between the fundamental stellar parameters determined with binary modelling, on the one hand, and asteroseismic scaling relations, on the other hand. 
} 

\subsubsection{Conclusions--Discussion}
{\citet{2018MNRAS.478.4669T} found agreement between the stellar parameters  derived with these two completely independent methods, but only when taking into account an empirically derived $\Delta\nu_{\rm ref}$-value. For KIC~8410637, KIC~5640750, and KIC~9540226, consistency between the asteroseismic and the dynamical stellar parameters required the usage of $\Delta\nu_{\rm ref}$ smaller than the usual solar reference value when the mass, $T_{\rm eff}$, the metallicity dependence as well as the surface effect were taken into account. They furthermore concluded that all three RGs are located on the RG branch, i.e., the RGs are H shell-burning stars. The importance of studying such systems is beyond doubt. They allow the determination of the fundamental stellar parameters, which are essential input of stellar models, in two ways. On the one hand, EBs offer a model-independent method to characterize the components by providing the global physical properties; on the other hand, a seismic study of the global $p$-mode pulsations of the RG star allows to unambiguously determine the same global properties, including also the evolutionary state. Such cases provide first-rate opportunities for {testing and understanding advanced} stellar evolution.
}
%
%
%
\section{Conclusions}

{{The majority of stars belong to a binary or a multiple system~\citep{2017PASA...34....1D}}. {We must therefore think 
of  a pulsating star in a binary or multiple system as the ``normal'' situation.}} It is clear that eclipsing systems with (a) 
pulsating component(s) are {essential and} challenging objects to study. {{Well detached, double-lined, 
eclipsing systems offer the advantage of model-independent fundamental parameters of their components which can be used as 
direct constraints in the search for relevant models. An additional constraint may come from the equal age and equal composition 
requirement. Whereas close eclipsing systems offer the possibility of studying the pulsations under the influence of other 
phenomena, e.g., equilibrium and dynamical tides, advanced evolution of (one of) the components (leading to angular momentum 
and mass transfer).}} To understand the {{past}} evolution of a system in terms of angular momentum and mass transfer, a long-term 
follow-up of the orbital parameters is a must, as demonstrated by the $\sim$10~yr study of the bright system RZ~Cas~\cite{2018MNRAS.475.4745M}. 
\par We can derive the principal characteristics of the pulsations from the residual light curves (sometimes also from high-resolution spectroscopy). 
However, we must also be able to trace their origin (i.e., find out which component pulsates in which modes~\cite{2012MNRAS.422..738S, 2014MNRAS.441.2515M}) 
and {{to reliably identify the modes}}. Since we can resolve the surfaces of pulsating stars in EBs during the eclipses, we can identify 
the spherical wave numbers $l$ and $m$ of the non-radial modes by exploiting the screening effect (using the spatial filtration or dynamic eclipse mapping method~\cite{2003ASPC..292..369G, 2005ASPC..333..258G, 2011MNRAS.416.1601B}) and a comparison can be performed 
between observed and predicted pulsation modes. {{The orbital properties too play a role, as we have seen, since different effects are detected in close, circular versus eccentric systems. In turn, tidally excited non-radial oscillations can affect the evolution of close binaries}
(for details see {\citet{2021RvMP...93a5001A}}, Section IV.F., and references therein).
{{Such in-depth studies are very pertinent and necessary for stellar physics in general. Not only for a deeper understanding of the pulsations under 
different circumstances (systems versus single stars), but also for a profound understanding of  the evolution of the rotational profile and angular momentum in stars and of the evolution of the systems, with as well as without tides.}}} Finally, we need advanced theoretical models to improve our understanding of 
the interior physics of the components themselves~\cite{2021RvMP...93a5001A}. {{Thanks to the excellent light curves acquired from space, we are just beginning to explore the complex reality of what could be called ``tidal asteroseismology''.}} 

\vspace{6pt}
\funding{This research received no external funding.} 

\institutionalreview{{Not applicable.}}

\informedconsent{{Not applicable.}}

\dataavailability{{Not applicable.}}

\acknowledgments{P.L. acknowledges useful discussions with her colleagues, C.~Aerts and \linebreak A.~Samadi~Ghadim, and apologizes for the (necessarily) subjective choice of the results presented in this review. {{Some on-going cited studies may have reached an advanced stage meanwhile.}} Many more fresh publications deal with this research topic, illustrating both the increased interest as well as the fast progress made in the field.}

\conflictsofinterest{The author declares no conflict of interest.} 

\newpage \abbreviations{Abbreviations}{The following abbreviations are used in this manuscript:\\

\noindent 
\begin{tabular}{@{}ll}
EA & Algol-type eclipsing binary \\
EB & eclipsing binary \\
MESA & Modules for Experiments in Stellar Astrophysics \\
oEA & oscillating Algol-type eclipsing binary \\
PHOEBE & PHysics Of Eclipsing BinariEs \\
RG & red giant \\
SB & spectroscopic binary \\
SD & semi-detached \\
TESS & Transiting Exoplanet Survey Satellite \\
ZAMS & Zero-Age Main Sequence \\
\end{tabular}}



\end{paracol}
\reftitle{References}





\end{document}